\documentclass[oupdraft]{bio2}
\usepackage{url}

\usepackage[french, english]{babel}
\usepackage{tikz}
\usepackage{helvet}
\usepackage{multirow}
\usepackage{amsmath}

\usepackage[figuresright]{rotating}
\usepackage{bbm}

\newcommand{\RN}[1]{
  \textup{\uppercase\expandafter{\romannumeral#1}}%
}

\def\bSig\mathbf{\Sigma}

\newcommand{\rev}[1]{#1}

\definecolor{green3}{RGB}{0,205,0}

\usepackage{subfig}
\usepackage{float}
\usepackage{graphicx}
\usepackage{xcolor}

\history{}

\begin{document}

\title{A joint model for multiple dynamic processes and clinical endpoints: application to Alzheimer's disease}

\author{C\'ecile Proust-Lima$^\ast$, Viviane Philipps, Jean-Fran\c cois Dartigues\\[4pt]
\textit{INSERM, UMR1219, Univ. Bordeaux, ISPED, Bordeaux, France}
\\[2pt]
{cecile.proust-lima@inserm.fr}}

\maketitle


\vspace*{1cm}
\noindent 
{\bf Abstract:} As other neurodegenerative diseases, Alzheimer's disease, the most frequent dementia in the elderly, is characterized by multiple progressive impairments in the brain structure and in clinical functions such as cognitive functioning and functional disability. Until recently, these components were mostly studied independently since no joint model for multivariate longitudinal data and time to event was available in the statistical community. Yet, these components are fundamentally inter-related in the degradation process towards dementia and should be analyzed together.
We thus propose a joint model to simultaneously describe the dynamics of multiple
correlated components. Each component, defined as a latent process, is measured by one or several continuous markers (not necessarily Gaussian). Rather than considering
the associated time to diagnosis as in standard joint models, we assume
diagnosis corresponds to the passing above a
covariate-specific threshold (to be estimated) of a pathological process which is modelled as a combination of the component-specific latent processes.
This definition captures the clinical complexity of diagnoses such as dementia diagnosis but also benefits from simplifications for the computation of Maximum Likelihood Estimates. We show that the model and estimation procedure can also handle competing clinical endpoints. The estimation procedure, implemented in a R package, is validated by simulations and the method is illustrated on a large French population-based cohort of cerebral aging in which we focused on the dynamics of three clinical manifestations and the associated risk of dementia and death before dementia.\\
{\bf Keywords:} aging; competing risks; dynamic model; joint model; latent process
model; multivariate longitudinal data\\



\maketitle


\section{Introduction} \label{sec1}

Dementia is a syndrome which mostly affects individuals 60 years and older, the most frequent type of dementia (approximately 60\%) being Alzheimer's disease. Dementia is characterized by a very long degradation process before clinical diagnosis which lasts a few decades (\citealp{amieva_prodromal_2008}). As other neuro-degenerative diseases, the degradation process in dementia has multiple anatomo-clinical components (\citealp{jack_tracking_2013}) including 
an accumulation of biomarkers in the brain ($\beta$-amyloid, $\tau$ protein), an alteration of the brain structure (atrophy of some regions such as the hippocampus), impaired clinical manifestations with the decline of several cognitive functions (e.g., executive functions, processing speed, episodic memory), an increase of functional limitations in the daily life (e.g. shopping, toileting, transferring from bed to chair), and possibly increased depressive symptoms among other behavioral alterations. 
Although inter-related, these components were mostly studied independently with a large focus on cognitive decline for which repeated measures have been available for long in cerebral aging cohorts (e.g., \citealp{proust-lima_joint_2016,graham_joint_2011}). 

The limitation to a single component (also called domain thereafter) was partly explained by the statistical complexity of analyzing multiple longitudinal components in link with a clinical endpoint. It requires the joint modelling of multivariate longitudinal markers and a survival process. A series of joint models for multiple longitudinal and survival data were proposed (see review in \citealp{hickey_joint_2016} and examples in \citealp{albert_approach_2010,andrinopoulou_joint_2014,baghfalaki_joint_2014,chi_joint_2006,choi_prediction_2014,lin_maximum_2002,pantazis_bivariate_2005}) but their application was limited to very simple cases. 
Indeed numerical complexity is considerable in joint models involving multiple longitudinal components due notably to an integral over the random effects in the likelihood which can not be solved analytically. The integral computation becomes cumbersome with more than a couple of random effects (\citealp{albert_approach_2010,rizopoulos_fully_2009}) and not necessarily accurately solved (\citealp{ferrer_joint_2016}). This entails that most applications focused on only two longitudinal markers (e.g. \citealp{lin_maximum_2002,li_joint_2012}) and/or random intercepts only (\citealp{li_joint_2012, tang_semiparametric_2015}) which is not sensible in complex diseases such as dementia. Contributions also mostly relied on a Bayesian estimation to circumvent the numerical issues (\citealp{andrinopoulou_joint_2014,baghfalaki_joint_2014,chi_joint_2006,choi_prediction_2014,tang_semiparametric_2015}) or a two-stage estimation (\citealp{albert_approach_2010}).

An additional problem in dementia as in many other diseases (especially in psychiatry and neurology) is that each domain is not directly measured; it is approached by multivariate markers. 
For example, cognitive level is usually measured by a battery of psychometric tests to apprehend all the coexisting cognitive functions, and brain structure comprises various regional volumes including the hippocampus volume but not limited to (\citealp{dickerson_cortical_2009}). Joint models for multivariate longitudinal markers measuring the same underlying process were proposed but were systematically limited to a univariate latent process (\citealp{he_joint_2016,luo_bayesian_2014}) such as cognition in dementia (\citealp{proust-lima_joint_2016}).
Finally, for psychometric and behavioral data at least, the usual assumption of normality does not hold and normalizing transformations have to be incorporated in the model to be able to rely on standard linear mixed regressions (\citealp{proust-lima_misuse_2011}).

In this context, our work aimed at developing a novel joint model for multivariate longitudinal markers measuring several latent processes and one or several clinical events. Instead of considering a standard proportional hazard model for the time to event, as mainly done in joint modelling framework, we considered 
a degradation process model for each clinical endpoint: the clinical endpoint is defined as a binary variable repeatedly collected at visits which becomes positive when its underlying continuous degradation process has passed above an unknown threshold. Using this definition, clinically relevant for dementia diagnosis and many other clinical endpoints, the joint model has an exact likelihood. As such, it can be applied to cases with more than just a couple of longitudinal markers and/or random effects. The replacement of the classical survival model by another model to provide exact likelihood had already been proposed in joint models with a univariate longitudinal Gaussian marker (\citealp{barrett_joint_2015}). In their work, the authors had opted for a sequential probit model defined in discrete time with which our approach has some similarities as explained in discussion. 

Section \ref{sec-model} describes the joint model for multiple latent domains and one clinical endpoint. Section \ref{sec-likelihood} details the likelihood computation and Section \ref{sec-extension} extends the model to other types of endpoints and competing endpoints. Section \ref{sec-simu} validates by simulations the estimation procedure which is implemented in a R package. Section \ref{sec-appli} details an application to three clinical manifestations in dementia, cognitive functioning, functional dependency and depressive symptomatology, in link with dementia diagnosis and in the presence of competing risk of dementia-free death. Finally, we conclude in Section \ref{sec-conclu}.

\section{The joint model for multiple latent domains and a clinical endpoint}
\label{sec-model}

\begin{figure}[!p]
\centering\includegraphics[width=0.9\textwidth]{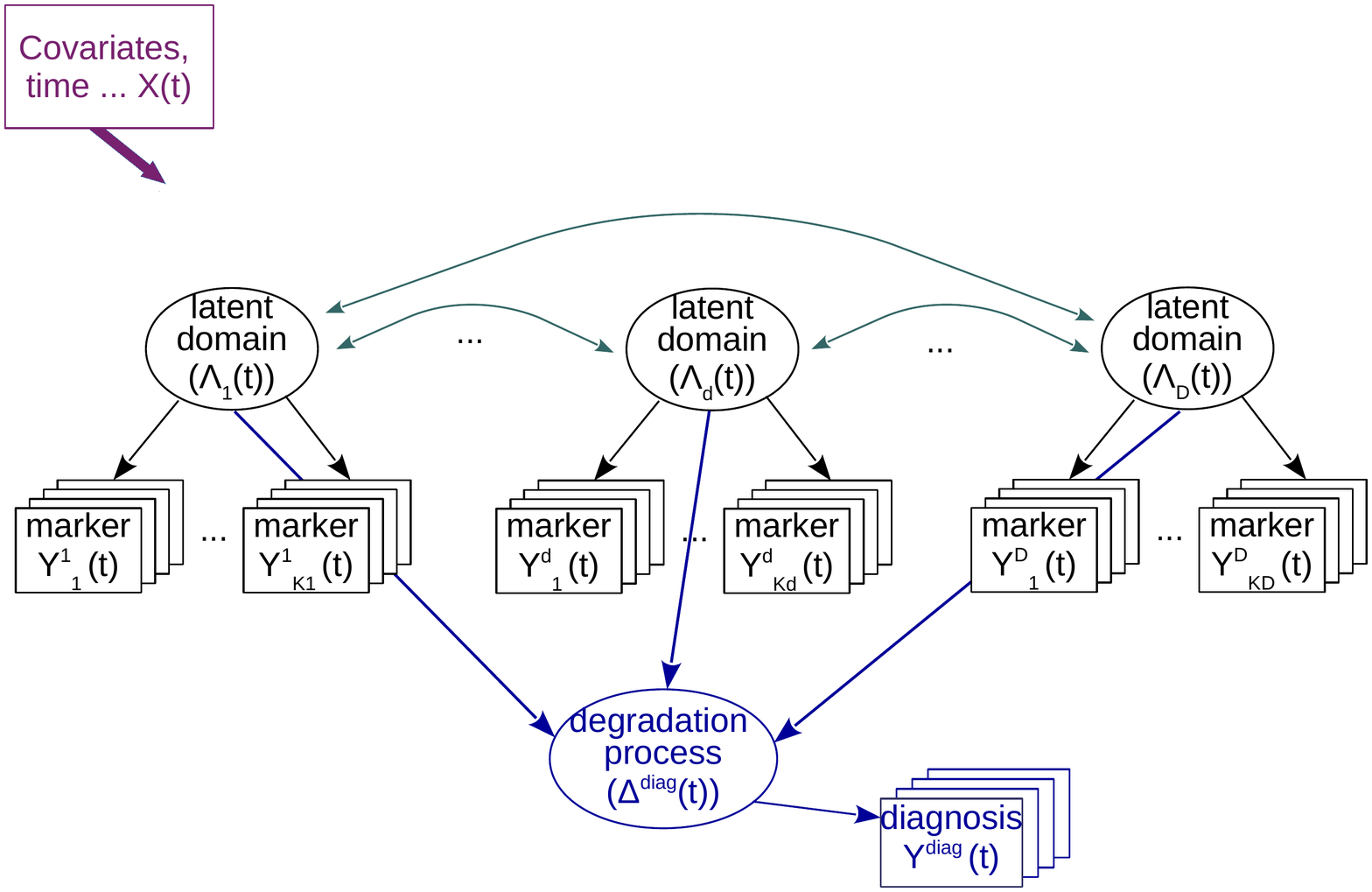}
\caption{Graph representing the joint model for $D$ latent domains noted $(\Lambda_d(t))$, each one measured repeatedly by several markers ($K_d$ for domain $d$) noted $Y^d_{k}$ and defining a global degradation process $(\Delta^{\text{diag}}(t))$ measured repeatedly by the diagnoses at follow-up visits $Y^{\text{diag}}$. 
For simplicity, subscript $i$ for individual $i$ is omitted.}
\label{fig_schema}
\end{figure}

The joint model for multiple latent domains and a clinical endpoint is described in Figure \ref{fig_schema} and formalized in Subsections \ref{sec_obs}, \ref{sec_LMM} and \ref{sec_degrad}.


\subsection{Latent domains measured by multivariate longitudinal markers}\label{sec_obs}

In a population of $N$ individuals, let consider $D$ latent domains (e.g., cognition, brain structure), each one defined for individual $i$ ($i=1,...,N$) as a latent process in continuous time $(\Lambda^d_{i}(t))$ with $t \in \mathbb{R}$ and $d=1,...,D$. Each latent domain $d$ is measured through a battery of $K_d$ markers repeatedly collected over time (e.g., several cognitive tests, volumes of different brain regions). Let define $Y^d_{kij}$ the measure of marker $k$ ($k=1,...K_d$) and individual $i$ for domain $d$ collected at time $t_{dkij}$ with $j=1,...n_{dki}$. 

To handle markers that are not necessarily Gaussian, we rely on previous works (\citealp{proust-lima_analysis_2013}) and assume that each marker, normalized by a parameterized link function, is a noisy measure of the underlying latent domain:

\begin{equation}\label{eqobs}
H^d_k(Y^d_{kij} ; \boldsymbol{\eta^d_k}) = \Lambda^d_i(t_{dkij}) + \epsilon^d_{kij}
\end{equation}

where $\epsilon^d_{kij}$ are centered independent Gaussian measurement errors with variance ${\sigma^d_k}^2$ and $H^d_k(.; \boldsymbol{\eta^d_k})$ is the link function which transforms the marker into a Gaussian framework. This link function depends on parameters $\boldsymbol{\eta^d_k}$ that are estimated on the data along with the other parameters. Any family of monotonically increasing parameterized transformations can be chosen for $H^d_k$, including the family of linear transformations which reduces to the standard Gaussian case. When departures from normality are suspected, we recommend the flexible and parsimonious family of linear combinations of quadratic I-splines $(\text{IS}_l)$ so that $H^d_k(x; \boldsymbol{\eta^d_k}) = \eta^d_{k0} + \sum_{l=1}^{n_{dk}+1} {\eta^d_{kl}}^2 \text{IS}_l(x)$ with $n_{dk}$ the number of knots kept small for parsimony (I-splines are integrated M-splines which ensure the monotonicity of the link functions, see \citealp{ramsay_monotone_1988}). \rev{Previous simulation studies demonstrated that using such nonlinear parameterized link functions yielded correct inference in mixed models when marker distribution deviated from normality (\citealp{proust-lima_misuse_2011})}. Note that although not detailed here for sake of brevity, marker-specific effects of covariates and/or random effects can be easily added to this equation of observation (\citealp{proust-lima_analysis_2013}). 

\subsection{Correlated trajectories of latent domains}\label{sec_LMM}

Each latent domain trajectory is described via a linear mixed model (\citealp{laird_random-effects_1982}):

\begin{equation}\label{Lambda}
\Lambda^d_i(t) = \boldsymbol{X_{id}(t)^\top \beta^d} + \boldsymbol{Z_{id}(t)^\top b^d_i} + w^d_{i}(t) ~~~, \forall t \in \mathbb{R}
\end{equation}

where $\boldsymbol{X_{id}(t)}$ is a vector of covariates associated with the vector of fixed effects $\boldsymbol{\beta^d}$ at the population level and $\boldsymbol{Z_{id}(t)}$ is a vector of covariates associated with the \rev{$q_d$-}vector of random effects $\boldsymbol{b^d_i}$ at the individual level with $\boldsymbol{b^d_i} \sim \mathcal{N}(\boldsymbol{0},\boldsymbol{B_{dd}})$. A zero-mean Gaussian process $w^d_{i}(t)$ can be added to make the trajectory more flexible at the individual level; for example a Brownian motion with covariance structure $\text{cov}(w^d_{i}(t),w^d_{i}(u)) = \sigma_{w^d}^2\text{min}(t,u)$ is often relevant in aging studies (\citealp{proust_nonlinear_2006,ganiayre_latent_2008}). We chose to keep the zero-mean Gaussian processes independent across domains. The correlation between the latent domains is only captured by correlations between the domain-specific random effects so that each individual is characterized by an overall vector of random effects:

\begin{equation}
\boldsymbol{b_i} = \left ( \begin{array}{l} \boldsymbol{b_i^1}\\ \vdots \\\boldsymbol{b_i^d}\\ \vdots \\ \boldsymbol{b_i^D}\\ \end{array}\right ) 
\sim  \mathcal{N}\left (
\left ( \begin{array}{l} 0\\ \vdots \\0\\ \vdots \\0\\ \end{array}\right ) ,
\boldsymbol{B}=\left ( \begin{array}{lllll} \boldsymbol{B_{11}} & ... & \boldsymbol{B_{1d}} & ... & \boldsymbol{B_{1D}} \\
\vdots & \ddots & \vdots & \ddots & \vdots\\
\boldsymbol{B_{1d}^\top} & ... & \boldsymbol{B_{dd}} & ... & \boldsymbol{ B_{dD}}\\
\vdots & \ddots & \vdots & \ddots & \vdots \\
 \boldsymbol{B_{1D}^\top} & ... & \boldsymbol{B_{dD}^\top} & ... & \boldsymbol{B_{DD}} \\
\end{array}\right ) \right )
\end{equation}

where $\boldsymbol{B_{de}}$ is the covariance matrix between random effects of latent domains $d$ and $e$ ($(d,e) \in \{1,D\}^2$). \rev{The unstructured $B$ matrix was parameterized through the $q=\sum_{d=1}^D q_d$ variances of random effects (parameter $\sigma.$ to estimate ($\sigma. \in \mathbb{R}$) for a variance of $\sigma.^2$) and the $\dfrac{q\times(q-1)}{2}$ correlations between pairs of random effects (parameter $\rho.$ to estimate ($\rho. \in \mathbb{R}$) for a correlation of $\dfrac{\exp(\rho.)-1}{1+\exp(\rho.)}$. }
As in any latent variable model, the dimension (location and scale) of the latent processes need to be defined to reach identifiability. This is achieved by standardizing each latent process $d$ with a zero mean intercept in $\boldsymbol{\beta^d}$ and a unit variance of the random intercept in $\boldsymbol{b^d_i}$. 

\subsection{Degradation process toward a clinical endpoint}\label{sec_degrad}

We assume there exists a degradation process in continuous time for each individual $i$ denoted $(\Delta^{\text{diag}}_i(t))$ with $t \in \mathbb{R}$. This process is measured by repeated observations of the clinical status of interest, dementia diagnosis in our case, $Y^{\text{diag}}_{ij}$ at time $t^{\text{diag}}_{ij}$ with $j$ the occasion ($j=1,...,n^{\text{diag}}_i$) so that an individual has the positive clinical status if the degradation process with an additional noise has reached a threshold (to estimate) at the visit time:
\begin{equation} \label{ydiag_def}
Y^{\text{diag}}_{ij} = 1 ~~~ \Leftrightarrow ~~~ \Delta^{\text{diag}}_i(t^\text{diag}_{ij}) + \epsilon^{\text{diag}}_{ij} ~ \geq ~ \zeta^{\text{diag}}_0 ~+ ~ \boldsymbol{{X^{\text{diag}}_{ij}}^\top \zeta^{\text{diag}}} 
\end{equation}

where $\epsilon^{\text{diag}}_{ij}$ is a zero-mean Gaussian independent random variable ($\epsilon^{\text{diag}}_{ij} \sim \mathcal{N}(0,1)$), $\zeta^{\text{diag}}_0$ is the threshold parameter in the reference group and $\boldsymbol{X^{\text{diag}}_{ij}}$ is a vector of covariates associated with parameters $\boldsymbol{\zeta^{\text{diag}}}$ that can modulate the threshold defining the positive clinical status. Note that observations are no more considered after a change to a positive clinical status. \rev{The variance of $\epsilon^{\text{diag}}_{ij}$ is constrained to 1 to ensure identifiability as shown in supplementary material, Section 1.1.}

The degradation process is defined as a linear combination of the latent domains: 
\begin{equation} \label{delta_def}
\Delta^{\text{diag}}_i(t) = \gamma_{1i}\Lambda^1_i (t) + ... + \gamma_{di} \Lambda^d_i (t) + ... + \gamma_{Di} \Lambda^D_i (t)
\end{equation}

where $\gamma_{di}$ is the intensity of latent domain $d$ contribution for individual $i$. It can be contrasted according to covariates with $\gamma_{di} =  \boldsymbol{{X^{\text{diag}}_{di}}^\top \gamma_{d}}$ in which $\boldsymbol{X^{\text{diag}}_{di}}$ is a vector of covariates (including the intercept) associated with the vector of parameters $\boldsymbol{\gamma_{d}}$. For instance, in dementia, the contributions of latent domains may be modulated by factors linked to cognitive reserve.

\section{Maximum Likelihood Estimation}
\label{sec-likelihood}

The entire vector of parameters denoted $\boldsymbol{\theta}$ comprises parameters $\{\boldsymbol{\eta_{k}^d},d=1,...,D,k=1,...,K_d \}$, $\{\boldsymbol{\beta^d},\sigma_{w^d}, \sigma^d_k, \boldsymbol{\gamma_{d}}, d=1,...,D\}$, $\zeta^{\text{diag}}_0$, $\boldsymbol{\zeta^{\text{diag}}}$ and all the parameters $\sigma$ and $\rho$ constituting the $\boldsymbol{B}$ matrix. It is estimated in the maximum likelihood framework. 

\subsection{Likelihood}\label{likelihood}

Let denote $\boldsymbol{Y^{\text{diag}}_i}= (Y^{\text{diag}}_{ij})_{j=1,...,n^{\text{diag}}_i}$ the repeated observed diagnoses, $\boldsymbol{Y_i}=(\boldsymbol{Y^d_i})_{d=1,...,D}$ the repeated and multivariate observed markers (with $\boldsymbol{Y^d_i}$ the vector of all the repeated and multivariate observations of latent domain $d$) and $\boldsymbol{\Lambda_i} = \left ( (\Lambda^d_i(t^{\text{diag}}_{ij}))_{j = 1,...,n^{\text{diag}}_i} \right )_{d=1,...,D }$ the set of latent processes at the observed diagnosis visits. \rev{By denoting $f(X)$ the generic probability density function of a random variable $X$}, the likelihood is $L(\boldsymbol{\theta}) = \displaystyle{\prod_{i=1}^N} f(\boldsymbol{Y^{\text{diag}}_i},\boldsymbol{Y_i}; \boldsymbol{\theta})$ with 
\begin{equation}
\begin{split}\label{jointdensity}
f(\boldsymbol{Y^{\text{diag}}_i},\boldsymbol{Y_i}) &= \int_{\mathbb{R}^{D n^{\text{diag}}_i}} f(\boldsymbol{Y^{\text{diag}}_i},\boldsymbol{Y_i}|\boldsymbol{\Lambda_i}) f(\boldsymbol{\Lambda_i})d\boldsymbol{\Lambda_i }= \int_{\mathbb{R}^{D n^{\text{diag}}_i}} f(\boldsymbol{Y^{\text{diag}}_i}|\boldsymbol{\Lambda_i})f(\boldsymbol{Y_i}|\boldsymbol{\Lambda_i})f(\boldsymbol{\Lambda_i})d\boldsymbol{\Lambda_i}\\
&= f(\boldsymbol{Y_i})  \int_{\mathbb{R}^{D n^{\text{diag}}_i}}  f(\boldsymbol{Y^{\text{diag}}_i}|\boldsymbol{\Lambda_i})  f(\boldsymbol{\Lambda_i}|\boldsymbol{Y_i})d\boldsymbol{\Lambda_i}
\end{split}
\end{equation}
where the vector of parameters $\boldsymbol{\theta}$ is omitted for simplicity and 
\begin{itemize}
\item the marginal density of the repeated markers $f(\boldsymbol{Y_i})$ can be decomposed into the multivariate normal density of the transformed observations of the markers through the link functions $\phi_{H(Y)}(\boldsymbol{H(Y_i)};\boldsymbol{\mu_{H(Y)}},\boldsymbol{V_{H(Y)}})$ times the Jacobian of the link functions $J(\boldsymbol{H(Y_i)})=\prod_{d=1}^D \prod_{k=1}^{K_d} \prod_{j=1}^{n_{dki}} J(H^d_k(Y^d_{kij} ; \boldsymbol{\eta^d_k}))$. The multivariate normal density of the transformed observations $\boldsymbol{H(Y)}$ has mean $\boldsymbol{\mu_{H(Y)}}=\boldsymbol{X_i \beta}$ and variance $\boldsymbol{V_{H(Y)}}= \boldsymbol{Z_iBZ_i}^\top + \boldsymbol{R_i} + \boldsymbol{\Sigma_i}$ with $\boldsymbol{\beta}= (\boldsymbol{\beta^d})_{d=1,...D}$, $\boldsymbol{X_i}$ and $\boldsymbol{Z_i}$ the D-block diagonal matrices with blocks $\boldsymbol{X^d_i}$ and $\boldsymbol{Z^d_i}$ that comprise row vectors $\boldsymbol{X_{id}(t)}^\top$ and $\boldsymbol{Z_{id}(t)}^\top$ for all the outcome-specific occasions $t$,  $\boldsymbol{R_i}$ the covariance matrix of the Gaussian processes $(w^d_i(t))$ at the same visits and $\boldsymbol{\Sigma_i}$ the diagonal variance matrix of measurement errors;

\item \rev{Following equations \eqref{ydiag_def} and \eqref{delta_def},} the conditional distribution of the repeated diagnoses is \rev{defined as follows (detailed calculations are provided in supplementary material, Section 1.2)}:

\begin{itemize}
\item $f(\boldsymbol{Y^{\text{diag}}_i}|\boldsymbol{\Lambda_i})=\Phi^{(n^{\text{diag}}_i)} \left (\boldsymbol{\zeta_i} - \boldsymbol{\Gamma_i \Lambda_i} ; \boldsymbol{0}, \boldsymbol{I_{n^{\text{diag}}_i}} \right )$ for subjects not diagnosed with dementia where $\Phi^{(n)}(\boldsymbol{x};\boldsymbol{m},\boldsymbol{V})$ is a $n$-dimensional Gaussian cumulative distribution function with mean $\boldsymbol{m}$ and variance $\boldsymbol{V}$ computed in $\boldsymbol{x}$, $\boldsymbol{\zeta_i}= \left ( \zeta^{\text{diag}}_0 ~+ ~ \boldsymbol{{X^{\text{diag}}_{ij}}}^\top \boldsymbol{\zeta^{\text{diag}}}\right )_{j = 1,...,n^{\text{diag}}_i}$, $\boldsymbol{\Gamma_i} = \left (\begin{array}{ccccc} \gamma_{1i} \boldsymbol{I_{n^{\text{diag}}_i}} &... &\gamma_{di} \boldsymbol{I_{n^{\text{diag}}_i}} &...&\gamma_{Di} \boldsymbol{I_{n^{\text{diag}}_i}} \end{array} \right )$ and $\boldsymbol{I_n}$ is the $n\times n$ identity matrix;
\item $f(\boldsymbol{Y^{\text{diag}}_i}|\boldsymbol{\Lambda_i})=\Phi^{(n^{\text{diag}}_i-1)}\left (\boldsymbol{\zeta_i^-} - \boldsymbol{\Gamma_i^- \Lambda_i} ; \boldsymbol{0}, \boldsymbol{I_{n^{\text{diag}}_i-1}} \right ) - \Phi^{(n^{\text{diag}}_i)}\left (\boldsymbol{\zeta_i} - \boldsymbol{\Gamma_i \Lambda_i} ;\boldsymbol{ 0}, \boldsymbol{I_{n^{\text{diag}}_i-1}} \right )$ for those diagnosed with dementia; $\boldsymbol{\zeta_i^-}$ and $\boldsymbol{\Gamma_i^-}$ are respectively $\boldsymbol{\zeta_i}$ and $\boldsymbol{\Gamma_i}$ without the last row for $t^{\text{diag}}_{in^{\text{diag}}_i}$;
\end{itemize}

\item the conditional density $f(\boldsymbol{\Lambda_i}|\boldsymbol{Y_i})$ is a multivariate normal density function $\phi_{\Lambda_i}(\boldsymbol{\Lambda_i};\boldsymbol{\mu_{\Lambda_i}},\boldsymbol{V_{\Lambda_i}})$ with mean $\boldsymbol{\mu_{\Lambda_i}}= \boldsymbol{\tilde{X}_i \beta} + \boldsymbol{V_{\Lambda H(Y)} {V_{H(Y)}}}^{-1} (\boldsymbol{H(Y_i)} -\boldsymbol{X_i \beta})$ and variance $\boldsymbol{V_{\Lambda_i}} = \boldsymbol{\tilde{Z}_iB\tilde{Z}_i}^{\top} + \boldsymbol{\tilde{R}_i} -
\boldsymbol{V_{\Lambda H(Y)} {V_{H(Y)}}}^{-1}\boldsymbol{{V_{\Lambda H(Y)}}}^\top$ where $\boldsymbol{\tilde{X}_i}$ and $\boldsymbol{\tilde{Z}_i}$ are the D-block diagonal matrices with blocks $\boldsymbol{\tilde{X}}_i^{d}$ and $\boldsymbol{\tilde{Z}}_i^{d}$ that comprise row vectors $\boldsymbol{X_{id}(t)}^\top$ and $\boldsymbol{Z_{id}(t)}^\top$ at diagnosis visits $t^{\text{diag}}_{ij}$ ($j=1,..., n^{\text{diag}}_i$), $\boldsymbol{\tilde{R}_i}$ the covariance matrix of the Gaussian processes $(w^d_i(t))$ at the same visits and $\boldsymbol{\tilde{\tilde{R}}_i}$ the covariance matrix of the Gaussian processes $(w^d_i(t))$ between diagnosis visits and marker visits so that $\boldsymbol{V_{\Lambda H(Y)}}= \boldsymbol{\tilde{Z}_i B Z_i}^\top + \boldsymbol{\tilde{\tilde{R}}_i}$.

\end{itemize}

The joint density of the observations can thus be rewritten:

\begin{equation}
\begin{split}\label{jointdensity2}
 f(\boldsymbol{Y^{\text{diag}}_i}&,\boldsymbol{Y_i}) =\phi_{H(Y)}(\boldsymbol{H(Y_i)};\boldsymbol{\mu_{H(Y)}},\boldsymbol{V_{H(Y)}})J(\boldsymbol{H(Y_i)}) ~~ \times \\
&  \begin{cases}
\displaystyle{\int_{\mathbb{R}^{Dn_i^{diag}}}} \Phi^{(n^{\text{diag}}_i)} \left (\boldsymbol{\zeta_i} - \boldsymbol{\Gamma_i \Lambda_i} ;\boldsymbol{ 0},\boldsymbol{ I_{n^{\text{diag}}_i}} \right )
\phi_{\Lambda_i}(\boldsymbol{\Lambda_i};\boldsymbol{\mu_{\Lambda_i}},\boldsymbol{V_{\Lambda_i}}) d\boldsymbol{\Lambda_i}& \text{if } Y^{\text{diag}}_{ij} = 0 ~\forall j \\
\displaystyle{\int_{\mathbb{R}^{Dn_i^{diag}}}} \left ( \Phi^{(n^{\text{diag}}_i-1)}\left (\boldsymbol{\zeta_i^-} - \boldsymbol{\Gamma_i^- \Lambda_i} ;\boldsymbol{ 0}, \boldsymbol{I_{n^{\text{diag}}_i-1}} \right ) \right. - & \\
$~~~~~~~~~~~~~~~~~$ \left. \Phi^{(n^{\text{diag}}_i)}\left (\boldsymbol{\zeta_i} - \boldsymbol{\Gamma_i \Lambda_i} ;\boldsymbol{ 0},\boldsymbol{ I_{n^{\text{diag}}_i}} \right ) \right )   \phi_{\Lambda_i}(\boldsymbol{\Lambda_i};\boldsymbol{\mu_{\Lambda_i}},\boldsymbol{V_{\Lambda_i}})d\boldsymbol{\Lambda_i}  & \text{if } Y^{\text{diag}}_{in^{\text{diag}}} = 1 \\
 \end{cases}
\end{split}
\end{equation}

Instead of approximating the multivariate integral over $\boldsymbol{\Lambda_i}$ by numerical integration techniques (as mostly done in joint modelling area with Gaussian quadratures), we use the properties of the skew-normal variables which provide a closed form for the integrals in equation \eqref{jointdensity2} (\citealp{arnold_flexible_2009}) (see Appendix for the general formula). The joint density becomes:

\begin{equation}
\begin{split}\label{jointdensity3}
f(\boldsymbol{Y^{\text{diag}}_i},\boldsymbol{Y_i}) &= \phi_{H(Y)}(\boldsymbol{H(Y_i)};\boldsymbol{\mu_{H(Y)}},\boldsymbol{V_{H(Y)}})J({\boldsymbol{H(Y_i)}}) ~~ \times \\
&  \begin{cases}
\Phi^{(n^{\text{diag}}_i)} \left (\boldsymbol{\zeta_i} - \boldsymbol{\Gamma_i \mu_{\Lambda_i}} ; \boldsymbol{0}, \boldsymbol{I_{n^{\text{diag}}_i}}  + \boldsymbol{\Gamma_i V_{\Lambda_i}  \Gamma_i} ^\top      \right )  & \text{if } Y^{\text{diag}}_{ij} = 0 ~\forall j \\
\Phi^{(n^{\text{diag}}_i-1)}\left (\boldsymbol{\zeta_i^-} - \boldsymbol{\Gamma_i^- \mu_{\Lambda_i}} ; \boldsymbol{0}, \boldsymbol{I_{n^{\text{diag}}_i-1}}  + \boldsymbol{\Gamma_i^- V_{\Lambda_i}  {\Gamma_i^-}}^{\top}      \right )  - & \\
$~~~~~~~~~~~~~~~~~$ \Phi^{(n^{\text{diag}}_i)} \left (\boldsymbol{\zeta_i} - \boldsymbol{\Gamma_i \mu_{\Lambda_i}} ; \boldsymbol{0}, \boldsymbol{I_{n^{\text{diag}}_i}}  + \boldsymbol{\Gamma_i V_{\Lambda_i}  \Gamma_i} ^\top      \right )  & \text{if } Y^{\text{diag}}_{in^{\text{diag}}} = 1 \\
 \end{cases}
\end{split}
\end{equation}

\subsection{Likelihood accounting for delayed entry}

When necessary, delayed entry can be accounted for by dividing the likelihood by the probability of being at risk of the clinical endpoint at study entry that is $Y^{\text{diag}}_{i1}=0$ at time $t^{\text{diag}}_{i1}$:
\begin{equation}\label{delayed}
L^{\text{delayed}}(\boldsymbol{\theta}) = \dfrac{L(\boldsymbol{\theta})}{\prod_{i=1}^N P(Y^{\text{diag}}_{i1}=0)}
\end{equation}
where 
\begin{equation}
\begin{split}
P(Y^{\text{diag}}_{i1}=0) &= \displaystyle{\int_{\mathbb{R}^{D}}} \Phi^{(1)} \left (\zeta_i^0 - \boldsymbol{\Gamma_i^0 \Lambda_i^0} ; 0, 1 \right )
\phi(\boldsymbol{\Lambda_i^0}; \boldsymbol{X_i^0 \beta},\boldsymbol{{Z_i^0 B Z_i^0}^\top}+\boldsymbol{R_i^0})d\boldsymbol{\Lambda_i^0} \\
&= \Phi^{(1)} \left ( \zeta_i^0 - \boldsymbol{\Gamma_i^0 X_i^0 \beta}; 0; 1 +  \boldsymbol{\Gamma_i^0 (Z_i^0 B {Z_i^0}^\top}+\boldsymbol{R_i^0}) \boldsymbol{{\Gamma_i^0}^\top} \right )
\end{split}
\end{equation}

with $\boldsymbol{\Lambda_i^0} = (\Lambda_{i}^d(t^{\text{diag}}_{i1}))_{d=1,...,D}$ the latent processes at entry, $\zeta_i^0 = \zeta^{\text{diag}}_0 + \boldsymbol{{X^{\text{diag}}_{i1}}^\top \zeta^{\text{diag}}}$ the threshold at entry, $\boldsymbol{\Gamma_i^0}=(\gamma_{1i},...,\gamma_{Di})$ the vector of domain-specific contributions, $\boldsymbol{X_i^0}$ and $\boldsymbol{Z_i^0}$ the D-block diagonal matrices with blocks $\boldsymbol{X^{d}_i(t^{\text{diag}}_{i1})}$ and $\boldsymbol{Z^{d}_i(t^{\text{diag}}_{i1})}$, and $\boldsymbol{R_i^0}$ the variance of the Gaussian processes at entry $w^d_{i}(t^{\text{diag}}_{i1})$ for $d=1,...,D$.

\subsection{Likelihood Optimization and implementation}

The log-likelihood is maximised using a modified Marquardt algorithm, a Newton-like algorithm, with convergence criteria based on parameter and log-likelihood stability and derivatives size  (\citealp{proust-lima_estimation_2017}). The latter is defined at iteration $l$ as $\frac{\nabla(L(\boldsymbol{\theta}^{(l)}))^\top\mathcal{H}^{(l)-1}\nabla(L(\boldsymbol{\theta}^{(l)}))}{n_\theta} \leq \omega $ with $\nabla(L(\boldsymbol{\theta}^{(l)}))$ and $\mathcal{H}^{(l)}$ the gradient and the Hessian matrix of the log-likelihood at iteration $l$, $n_\theta$ the length of $\boldsymbol{\theta}$, and $\omega$ the convergence threshold (fixed here at 0.001). This criterion is very stringent and ensures convergence toward a maximum. The program was implemented in C++ with an interface in R and parallel computations to fasten the estimation procedure. The R package called \texttt{multLPM} can be downloaded on github: (\url{https://github.com/VivianePhilipps/multLPM}). \rev{The program fits any model that has the same structure as defined in Section \ref{sec-model} (and possibly a second competing event as defined in Section \ref{sec-extension})}. The multivariate normal cumulative distribution functions were computed by using Genz routines (\citealp{genz_numerical_1992}).

\section{Extension to multiple clinical endpoints and continuous-time events}
\label{sec-extension}
\subsection{Multiple endpoints at predefined visits}

The definition of the degradation process and measurement model for a clinical event observed at specific visits such as diagnoses (Subsection \ref{sec_degrad}) extends naturally to the case of two causes of diagnosis (with cause-specific degradation processes). The log-likelihood has the exact same structure except that the dimension of the cumulative distribution functions is augmented to the number of visits for each cause of clinical event, $2\times n^{\text{diag}}_i$ if the two diagnoses are made at the same times. 

\subsection{Event in continuous time}\label{sec_degrad_death}

The model definition relies on discrete repeated visits for the clinical event and as such does not directly extend to clinical events that are intrinsically defined in continuous time such as death. With events in continuous time, a preliminary discretization of time is necessary: we partition time into $S$ intervals $\mathcal{I}_s=[l_s,u_s]$ with midpoints $m_s$ for $s=1,...,S$ ($l_s=u_{s-1}$ for $s>1$). A subject is followed from interval $s_{0i}$, the interval containing the exact entry time in the study, and is followed until interval $s_i$: if the subject has the event, $s_i$ is such that the time of event is in $\mathcal{I}_{s_i}$; if the subject is censored, $s_i$ is such that the censoring time is in $\mathcal{I}_{s_i+1}$. Without loss of generality, we focus on death and define the repeated death status $Y^{\text{death}}_{ij}$ in each interval $j$ ($j=s_{0i},...,{s_{i}}$) with $Y^{\text{death}}_{ij}=0$ for all $j$ except $j=s_i$ for those who die ($Y^{\text{death}}_{is_i}=1$). 

%
%
%

Using the exact same definition as for clinical diagnoses, we consider a latent degradation process defined as a linear combination of the latent domains:
\begin{equation}
\Delta^{\text{death}}_i(t) = \delta_{1i}\Lambda^1_i (t) + ... + \delta_{di} \Lambda^d_i (t) + ... + \delta_{Di} \Lambda^D_i (t)
\end{equation}

We then define the probability of dying in interval $s$ as the probability that the noisy underlying degradation process is above a specific threshold at the midpoint of interval $s$:
\begin{equation}
P(Y^{\text{death}}_{is} = 1) = P(\Delta^{\text{death}}_i(m_s) + \epsilon^{\text{death}}_{is} ~ \geq ~ \zeta^{\text{death}}_0 ~+ ~ \boldsymbol{{X^{\text{death}}_{is}}^\top \zeta^{\text{death}}})
\end{equation}
As previously, $\delta_{di}$ characterizes the intensity of latent domain $d$ contribution for individual $i$ and it can be contrasted according to covariates with $\delta_{di} = \boldsymbol{{X^{\text{death}}_{di}}^\top \delta_{d}}$ in which $\boldsymbol{X^{\text{death}}_{di}}$ is a vector of covariates (including the intercept) associated with the vector of parameters $\boldsymbol{\delta_{d}}$; $\epsilon^{\text{death}}_{is}$ is a zero-mean Gaussian independent random variable ($\epsilon^{\text{death}}_{is} \sim \mathcal{N}(0,1)$); $\zeta^{\text{death}}_0$ is the threshold in the reference group and $\boldsymbol{X^{\text{death}}_{is}}$ is a vector of covariates associated with parameters $\boldsymbol{\zeta^{\text{death}}}$ that can modulate the threshold defining the death status. In particular, $\boldsymbol{X^{\text{death}}_{is}}$ may include time. 

When this discretized event replaces clinical diagnosis, the likelihood has the same structure except that the dimension of the cumulative distribution function becomes $s_{i}-s_{0i}+1$. In case of delayed entry, the likelihood is divided by the probability to have survived until $l_{s_{0i}}$.

\subsection{Endpoint at predefined visits in competition with an event in continuous time}\label{compet}

Without loss of generality, we take the example of a clinical diagnosis for which death before diagnosis constitutes a competing event. The methodology handles this by combining the two degradation processes of sections \ref{sec_degrad} and \ref{sec_degrad_death}, and by slightly changing the observations for death. To focus on death before diagnosis, we only consider death in the $x$ years following a negative diagnosis ($x$=3 years is realistic in the case of dementia). Otherwise, time to death is censored at the last visit $t_{in_i^{\text{diag}}}^{\text{diag}}$. 
With these changes, the joint log-likelihood has again the exact same structure as in section \ref{likelihood} except that the dimension of the cumulative distribution functions is augmented to the number of visits for the clinical event and the number of intervals for the discretized event $n^{\text{diag}}_i + s_{i}-s_{0i}+1$. In case of delayed entry, the likelihood is divided by the probability to have survived until $s_{0i}-1$ and still be at risk of clinical event at entry.

\section{Numerical evaluation of the methodology}
\label{sec-simu}

\rev{We evaluated the methodolgy with two series of simulations. We first validated the estimation procedure under a correctly specified model by investigating different combinations of key simulation parameters (i.e., number of subjects, proportion of events, nature of the event and uncertainty in the markers measurement). We then evaluated the behavior of the method under different types misspecification (i.e., distribution of errors in the degradation model, correlation between the latent domains, second clinical endpoint or additional domain not modelled). For all scenarios, design and parameter values were inspired by the application data, and 200 replications were done. All the scenarios are summarized in supplementary Table S1.}

\subsection{Simulation study I: Validation of the estimation procedure}


\rev{We considered two longitudinal domains, namely cognition and functional dependency, measured by two psychometric tests and one scale, respectively. We generated linear trajectories of the two domains according to age and adjusted for binary educational level (0.5-probability Bernoulli) and their interaction for domain 1. Correlated random intercept and slope with age took into account the correlation within and between domains at the individual level. Entry in the cohort was generated from a normal distribution (mean 75, standard deviation 3). Observed visits were generated with a uniform distribution in [-1,1] years around the theoretical visits every 2.5 years up to 20 years, and a 15\% dropout was considered at each visit. We primarily considered that the two domains contributed to the clinical event model and that a clinical diagnosis was made at each visit. Longitudinal data were censored after the event. We considered Gaussian longitudinal markers with different ranges inspired by psychometric tests IST and DSST for cognition and IADL sum-score for function (see application Section \ref{sec-appli}). More details on the generating procedure are given in supplementary section 2.1}.

\rev{In the main scenario (I.1.a), generating parameters roughly corresponded to those obtained on the application data and samples included 500 subjects. Over 200 replicates, this lead on average to 4.3 visits per subject and 150 diagnoses (30\%). As reported in Table \ref{tab_simu1}, parameters were well estimated without bias and no departure from the expected 95\% coverage rate of the 95\% confidence interval. In additional simulations, we also considered: (scenario I.1.b) smaller samples of 200 subjects which lead in mean to 59 (29.5\%) diagnoses; (scenarios I.2.a and I.2.b) smaller proportion of diagnoses by changing the values of each domain contribution. This lead on average to 56 (11\%) and 22 (11\%) diagnoses in samples of 500 and 200 subjects, respectively; (scenario I.3) larger measurement errors for the markers; (scenario I.4) continuous event discretized into 10 intervals instead of a clinical diagnosis made at each visit. In all these scenarios, inference remained correct with negligible bias and satisfying coverate rates (see supplementary Tables S2 to S6). With 200 subjects, variances were systematically higher though. Overall, these simulations validated the estimation procedure. }

\begin{table}[!p]
\caption{Simulation results for 2 longitudinal domains and diagnosis of dementia on 200 replicates of 500 subjects (\rev{Scenario I.1.a -} 194 models converged in less than 30 iterations). $\theta$ refers to the generated value and $\overline{\hat{\theta}}$ to the mean estimate over the replicates. The bias is reported relative to the expected value in \%. SD$(\hat{\theta})$ refers to the empirical standard deviation of the estimates, $\overline{\widehat{\text{SE}(\hat{\theta})}}$ refers to the mean standard error of the estimates, and 95\%CR refers to the coverage rate of the 95\% confidence interval of the estimate.} \label{tab_simu1}
\begin{tabular}{lrrrrrr}
\hline
Parameter    & $\theta$ & $\overline{\hat{\theta}}$ & bias (\%) & $\overline{\widehat{\text{SE}(\hat{\theta})}}$ & SD$(\hat{\theta})$ & 95\%CR\\
\hline
\multicolumn{7}{l}{ \emph{Model for Domain 1 (markers m1 and m2)} } \\
  Intercept$^*$                                &   0 & 0 & - &-  & - &-  \\                      
age                                            &   1.000 & 1.004 & 0.4 & 0.066 & 0.061 & 94.8 \\          
 EL                                            &   -1.100 & -1.105 & 0.4 & 0.118 & 0.122 & 94.3 \\   
age$\times$EL                                  &   0.100 & 0.102 & 1.9 & 0.058 & 0.057 & 94.3 \\     
Transformation parameter 1 (m1)                &   18.000 & 17.996 & $<$0.1 & 0.323 & 0.329 & 94.3 \\      
Transformation parameter 2 (m1)                &   4.000 & 3.985 & 0.4 & 0.189 & 0.194 & 93.3 \\     
SD of error (m1)                               &   0.800 & 0.805 & 0.6 & 0.041 & 0.042 & 94.8 \\      
Transformation parameter 1 (m2)                &   20.000 & 20.004 & $<$0.1 & 0.390 & 0.397 & 95.9 \\     
Transformation parameter 2 (m2)                &   5.000 & 4.979 & 0.4 & 0.226 & 0.226 & 94.8 \\     
SD of error (m2)                               &   0.300 & 0.302 & 0.5 & 0.015 & 0.015 & 95.9 \\ 
\multicolumn{7}{l}{ \emph{Model for Domain 2 (marker m3)} }   \\         
Intercept$^*$                                  &   0 & 0 &-  &-  &-  &-  \\                      
Age                                            &   1.700 & 1.730 & 1.8 & 0.205 & 0.202 & 96.9 \\         
 EL                                            &   -0.500 & -0.510 & 2.0 & 0.123 & 0.127 & 95.4 \\      
Transformation parameter 1 (m3)                &   5.000 & 5.001 & $<$0.1 & 0.010 & 0.010 & 97.9 \\        
Transformation parameter 2 (m3)                &   0.100 & 0.099 & 0.9 & 0.011 & 0.010 & 97.4 \\     
SD of error (m3)                               &   0.900 & 0.914 & 1.6 & 0.106 & 0.104 & 96.4 \\ 
 \multicolumn{7}{l}{ \emph{Variance Covariance matrix of random effects for Domain 1 (d1) and Domain 2 (d2)}}  \\        
SD intercept (d1)$^*$                          &   1 & 1 & - &-  &-  &-  \\                      
SD slope (d1)                                  &   0.513 & 0.515 & 0.5 & 0.032 & 0.031 & 95.4 \\          
SD intercept (d2)$^*$                          &   1 & 1 & - & - &-  &-  \\                      
SD slope (d2)                                  &   0.883 & 0.891 & 0.9 & 0.086 & 0.090 & 95.4 \\        
Corr$^\dag$ intercept (d1) and slope (d1)      &   -0.833 & -0.829 & 0.5 & 0.137 & 0.140 & 94.3 \\   
Corr$^\dag$ intercept (d2) and slope (d2)      &   -1.070 & -1.048 & 2.1 & 0.232 & 0.225 & 95.9 \\   
Corr$^\dag$  intercept (d1) and intercept (d2) &   0.905 & 0.936 & 3.4 & 0.223 & 0.231 & 97.9 \\         
Corr$^\dag$  slope (d1) and intercept (d2)     &   -0.956 & -0.970 & 1.4 & 0.253 & 0.262 & 98.5 \\     
Corr$^\dag$  intercept (d1) and slope (d2)     &   -0.297 & -0.308 & 3.6 & 0.173 & 0.173 & 95.9 \\   
Corr$^\dag$  slope (d1) slope (d2)             &   1.899 & 1.952 & 2.8 & 0.265 & 0.259 & 95.9 \\   
\multicolumn{7}{l}{  \emph{Model for the clinical endpoint}}  \\  
Threshold                                      &   3.000 & 3.021 & 0.7 & 0.180 & 0.178 & 96.9 \\     
Contribution of domain 1                       &   0.300 & 0.302 & 0.5 & 0.061 & 0.065 & 92.8 \\     
Contribution of domain 2                       &   0.400 & 0.402 & 0.4 & 0.066 & 0.064 & 96.4 \\     
\hline
\end{tabular}

\vspace*{1cm}
$^*$ fixed parameter; $^\dag$ transformed correlation parameter
\end{table}

\subsection{Second simulation study: Behavior under misspecification}

\rev{Based on the same overall design of simulations, we investigated 4 types of misspecification. Results are summarized in supplementary material, section 2. When generating logistic errors in the degradation process toward diagnosis (in equation \eqref{ydiag_def}) instead of Gaussian errors (supplementary Table S7) or when neglecting the intra-individual correlation between latent domains (supplementary Table S8), inference quality was not much affected. When a second event was not modelled (supplementary Table S9), results depended on the level of association between domains and neglected competing event. Inference was not affected with very small contributions of the domains but as expected, parameters of the degradation process model became slightly biased with increased contributions (as it lead to informative censoring). When a third correlated latent domain was not modelled, the inference at the latent domain (and marker) level remained correct. The estimates of the degradation process differed depending on the intensity of association with the neglected domain but this was expected as the interpretation of the contributions (adjusted or not for other domains) differ (supplementary Table S10).
}

\section{Application to clinical manifestations in dementia}
\label{sec-appli}

Changes in various clinical measures such as cognitive tests, dependency scales or depressive symptomatology have been separately observed in prodromal dementia (\citealp{amieva_prodromal_2008}) suggesting a possible concomitant role in the dementia process with a modulation of the intensity by educational level which illustrates a potential compensatory mechanism. Our objective was to precisely investigate the role of cognition, dependency and depression in the degradation process toward dementia by jointly analyzing their trajectories and their determinants in link with dementia diagnosis and accounting for the competing death. 

\subsection{The PAQUID data}
We relied on the data from the population-based PAQUID cohort which included 3777 individuals aged 65 years and older and living at home in southwestern France in 1989-1990. Individuals were then followed for up to 25 years with repeated neuropsychological evaluations and clinical diagnoses of dementia every 2 or 3 years (\citealp{letenneur_incidence_1994}) and death continuously recorded. We focused on a subsample of 646 individuals who were tested for ApoE4, the main genetic factor associated with aging. We excluded 22 individuals diagnosed with dementia at baseline and 31 individuals who did not have at least one observation of each clinical measure. The final sample consisted of 593 individuals among which 180 developed a dementia and 283 died in the three years following a negative dementia diagnosis. The sample comprised 332 (56\%) women, 438 (74\%) individuals with higher education level (EL+; individuals who graduated from primary school) and 130 (22\%) carriers of at least one copy of the APOE4 allele. The mean age at entry was 73.6 (SD=6.1) years.

The three clinical manifestations were: 
\begin{itemize}
\item \emph{Cognitive impairment}; it was assessed by four psychometric tests (inverted so that higher levels indicated higher impairment). The Mini-Mental State Examination (MMSE) provides an index of global cognitive performance, the Benton Visual Retention Test (BVRT) assesses visual memory, the Isaacs Set Test (IST) measures verbal semantic memory and speed, and the Digit Symbol Substitution Test (DSST) provides a global measure of executive functioning and processing speed.
\item \emph{Functional dependency}; it was assessed by the French version of the Instrumental Activities of Daily Living (IADL). We summed the grades of dependency for four activities, telephone use, transportation, medication and domestic finances.
\item \emph{Depressive symptomatology}; it was assessed by the sum-score of the Center for Epidemiologic Studies Depression Scale (CES-D).
\end{itemize}

Individuals had between 1 and 12 repeated measures of each marker with a median of 6 (Interquartile range IQR=4-8) for MMSE, IADL and dementia diagnosis, 5 (IQR=4-8) for CES-D, 5 (IQR=3-8) for BVRT, 4 (IQR=3-7) for IST and 4 (IQR=2-6) for DSST. 

\subsection{Model specification}

The structure of the model is summarized in Figure \ref{fig_schema_appli} and specifics are given below.

\begin{figure}[!p]
\centering\includegraphics[width=0.9\textwidth]{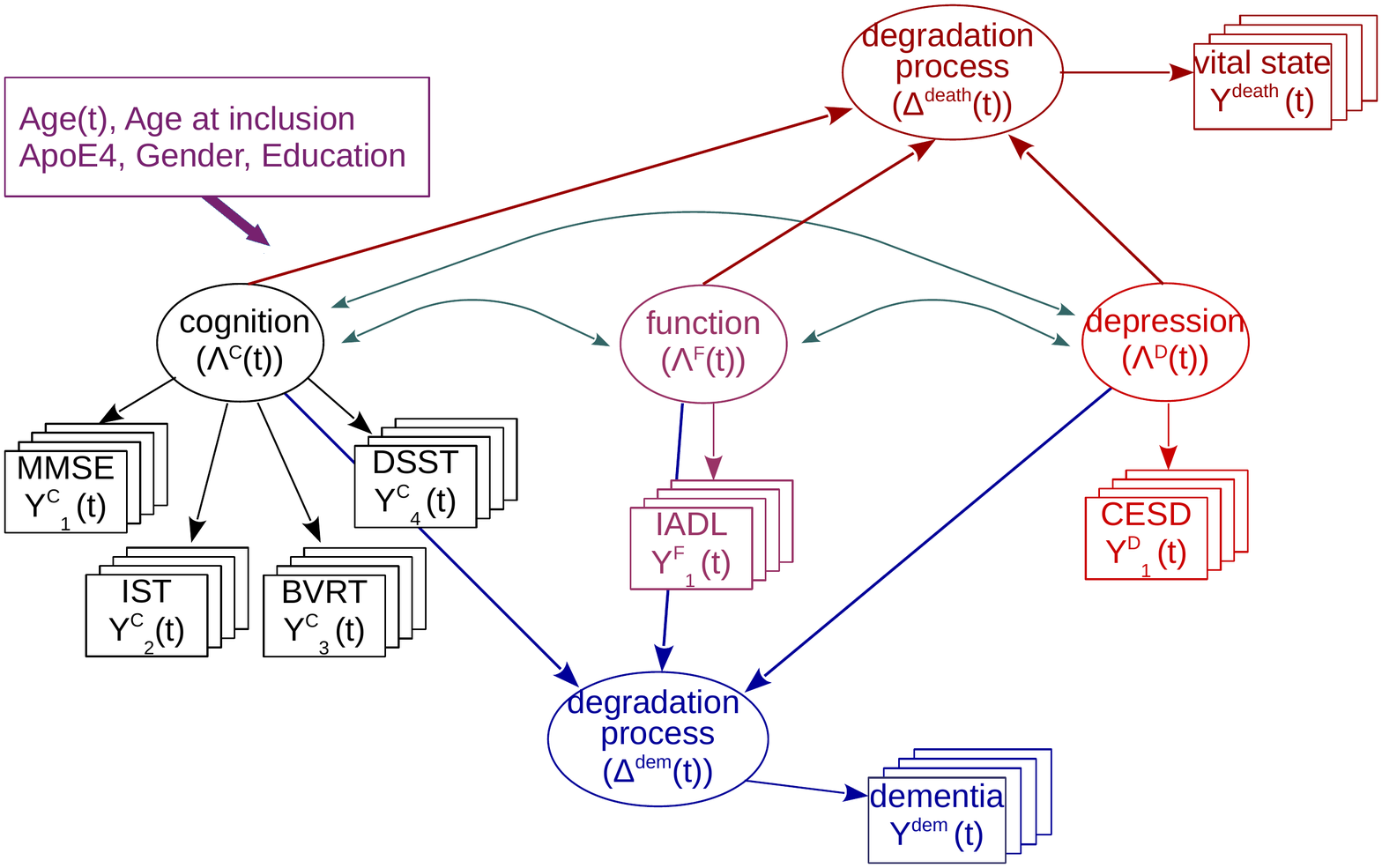}
\caption{Graph representing the joint model applied to PAQUID data with 3 latent domains, cognition, function and depression, measured respectively by MMSE, IST, BVRT and DSST cognitive scores, IADL functional scale and CES-D scale of depressive symptoms. Latent domains define two global degradation processes toward dementia and death before dementia. For simplicity, subscript $i$ for individual $i$ is omitted.}
\label{fig_schema_appli}
\end{figure}

\emph{Quadratic trajectories according to age} were assumed for each domain with an adjustment for age at entry, EL+, ApoE4 and gender (and their interactions with age and age squared), and three individual random effects (on intercept, slope and slope squared) correlated within and between domains. An additional time dependent variable indicating the baseline evaluation was included due to evidence of a primo passation negative effect. The selection of covariates and interactions was determined in separate analyses by domain with a 20\% significance level. 

\rev{\emph{Marker-specific link functions} used in equation \eqref{eqobs} to normalize the markers were quadratic I-splines functions with 3 internal knots placed at the quartiles of the marker distribution over follow-ups}. The relevance of the splines transformations was checked in domain-specific analyses by comparing the Akaike criterion (AIC) with the AIC of the model assuming a linear link. 

\emph{Degradation process toward dementia} was modelled according to the three domains with an adjustment for age at entry, gender, ApoE4 carriers and EL+ and a potential change in the threshold of dementia after the 10-year follow-up visit (due to a new drug on the market that implied earlier diagnoses). 

\emph{Degradation process toward dementia-free death} was also modelled according to the three domains. We discretized death into 8 intervals with boundaries at 65, 70, 74, 77, 80, 83, 86, 90 and 104 years chosen according to the distribution of death times. The threshold defining the probability of dying was modelled according to age using natural cubic B-splines with knots at 70, 83, 90 and 100 years.

As for any complex model, we recommend to estimate the model progressively by first fitting domain-specific mixed models separately then jointly and finally with dementia and death models. This provides at each step reasonable initial values and reduces computational time. 

\subsection{Results}

Fixed effects of the multivariate mixed model for cognition, functional dependency and depressive symptoms are given in Table \ref{tab_estimates} and predicted trajectories by covariate profile are displayed in Figure S1 of Supplementary Materials. In summary, each domain had a quadratic trajectory with age characterized by an acceleration in older ages. Individuals included at an older age were systematically more impaired than those included at an younger age (whatever the current age) highlighting a cohort effect. The first passing indicator confirmed that evaluations at baseline underestimated cognitive level and depressive symptoms while overestimating functional dependency. More educated individuals had better cognitive and functional levels. ApoE4 carriers had a faster increase of cognitive and functional impairments. Finally, men had a lower level of depressive symptoms at 65 years but the difference with women reduced when age increased. Men also had a slower increase of functional dependency.

\begin{table}[!p]
\caption{Fixed effect estimates of multivariate mixed submodel on PAQUID data (N=593) with three domains (cognition, functional dependency and depressive symptoms). } \label{tab_estimates}
\centering
\begin{tabular}{lrrrrrrrrr}
\hline
& \multicolumn{3}{c}{\emph{Cognition}} &
                                         \multicolumn{3}{c}{\emph{Functional dependency}}& \multicolumn{3}{c}{\emph{Depressive symptoms}} \\
Covariate               & $\theta$ & SE    & p-value$^{\ddag}$ & $\theta$ &
                                                                      SE    & p-value  & $\theta$ & SE    & p-value\\
\hline
Intercept$^*$ & 0 & - & - & 0 & - & -&0 & -  & - \\
Age  $^{\dag}$                 & -0.306   & 0.141 & 0.030    & -1.371   & 0.323 & $<$0.001 & 0.189    & 0.162 & 0.244    \\
  Age$^2$     $^{\dag}$          & 0.488    & 0.066 & $<$0.001 & 1.162    & 0.186 & $<$0.001 & 0.053    & 0.053 & 0.322    \\
\multicolumn{3}{l}{{\small \emph{Test for (Age, Age$^2$)}}} & {\small  \emph{ $<$0.001}} &&&{\small  \emph{ $<$0.001 }} &&&{\small  \emph{ $<$0.001 }} \\
  Initial age  $^{\dag}$         & 0.372    & 0.102 & $<$0.001 & 0.211    & 0.160 & 0.188    & -0.173   & 0.075 & 0.021    \\
  First passing effect  & 0.184    & 0.035 & $<$0.001 & -0.257   & 0.071 & $<$0.001 & 0.506    & 0.065 & $<$0.001 \\
EL+                     & -1.543   & 0.196 & $<$0.001 & -0.614   & 0.160 & $<$0.001 &          &       &          \\
EL+ $\times$ Age        & 0.010    & 0.101 & 0.922    &          &       &          &          &       &          \\
EL+ $\times$ Age$^2$    & 0.044    & 0.041 & 0.282    &          &       &          &          &       &          \\
\multicolumn{3}{l}{{\small \emph{Test for EL+ $\times$ (Age, Age$^2$)}}} &{\small  \emph{ 0.182 }} &&&&&&\\
ApoE4+                  & -0.136   & 0.156 & 0.385    & -0.030   & 0.260 & 0.908    & & &                              \\
ApoE4+ $\times$ Age     & 0.221    & 0.214 & 0.303    & -0.554   & 0.462 & 0.230         & & &                         \\
ApoE4+ $\times$ Age$^2$ & 0.063    & 0.085 & 0.462    & 0.494    & 0.186 & 0.008         & & &                         \\
\multicolumn{3}{l}{{\small \emph{Test for ApoE4+ $\times$ (Age, Age$^2$)}}} & {\small \emph{  $<$0.001 }} &&& {\small \emph{  $<$0.001} } &&&\\
Male                    &          &       &          & 0.076    & 0.194 & 0.694    & -0.859   & 0.176 & $<$0.001 \\
Male $\times$ Age       &          &       &          & -0.184   & 0.320 & 0.565    & 0.108    & 0.216 & 0.616    \\
Male $\times$ Age$ ^2$  &          &       &          & -0.045   & 0.114 & 0.695    & 0.051    & 0.072 & 0.476    \\
\multicolumn{3}{l}{{\small \emph{Test for Male $\times$ (Age, Age$^2$)}}} &  &&& {\small \emph{ 0.038}} &&& {\small  \emph{ $<$0.001 } } \\
\hline
\end{tabular}
\newline
$^*$ fixed parameter; $^{\dag}$ Age and Initial age are indicated in decades and centered around 65 years old; $^{\ddag}$ p-value of the univariate Wald test (or bivariate for italic lines); 
\end{table}

\begin{table}[!p]
\caption{\rev{Fixed effect estimates of degradation process submodels for dementia and dementia-free death on PAQUID data (N=593) defined either as a combination of the three domains (cognition, functional dependency and depressive symptoms) (left panel) or from a single domain in separated univariate models (righ panels). Are also reported conditional information criteria for dementia and death data conditional on longitudinal information (conditional AIC, conditional BIC, UACV)}.} \label{tab_estimates2}
\setlength{\tabcolsep}{5pt}
\centering
{\footnotesize
\begin{tabular}{llrrrrrrrrrrrr}
\hline
& & \multicolumn{3}{c}{Three domains}& \multicolumn{9}{c}{Single domain}  \\
\cline{6-14}
& & \multicolumn{3}{c}{simultaneously}& \multicolumn{3}{c}{Cognition} & \multicolumn{3}{c}{Dependency} & \multicolumn{3}{c}{Dep. symptoms}  \\
\multicolumn{2}{l}{Covariate} & $\theta$ & SE & p & $\theta$ & SE  & p & $\theta$ & SE  & p & $\theta$ & SE  & p \\
\hline
\multicolumn{5}{l}{  \emph{Degradation process toward dementia}}\\

  Threshold    & Intercept    & 2.744   & 0.268   & $<$0.001 & 2.885  & 0.239 & $<$0.001 & 2.706  & 0.066 &$<$0.001 & 2.515  & 0.074 &$<$0.001 \\ 
               & EL+          & -0.285  & 0.156   & 0.067    & -0.642 & 0.131 & $<$0.001 & 0.016  & 0.100 &0.873    & 0.285  & 0.096 &0.003 \\ 
               & ApoE4+       & -0.008  & 0.151   & 0.957    & -0.344 & 0.118 & 0.003    & -0.088 & 0.209 &0.674    & -0.345 & 0.092 &$<$0.001 \\ 
               & Male         & 0.056   & 0.133   & 0.674    & 0.176  & 0.108 & 0.102    & -0.129 & 0.103 &0.212    & 0.029  & 0.090 &0.750 \\ 
               & Init. age$^*$ & 0.198  & 0.119   & 0.097    & 0.008  & 0.011 & 0.486    & -0.010 & 0.009 &0.245   & -0.062 & 0.006 &$<$0.001 \\ 
               & 10y visit    & -0.418  & 0.134   & 0.002    & -0.605 & 0.125 & $<$0.001 & -0.556 & 0.103 &$<$0.001 & -0.972 & 0.082 &$<$0.001 \\ 
  Cognition    & Intercept    & 0.511   & 0.086   & $<$0.001 & 0.706  & 0.064 & $<$0.001 &        &       &        &        &       & \\ 
  Dependency   & Intercept    & 0.273   & 0.060   & $<$0.001 &        &       &          & 0.313  & 0.066 &$<$0.001&        &       & \\
               & ApoE4+       & 0.136   & 0.069   & 0.048    &        &       &          & 0.097  & 0.062 &0.119   &        &       & \\ 
  Dep. Sympt.  & Intercept    & -0.447  & 0.147   & 0.002    &        &       &          &        &       &        & 0.036  & 0.089 & 0.687 \\ 
               & EL+          & 0.452   & 0.160   & 0.005    &        &       &          &        &       &        & 0.246  & 0.108 &0.023 \\ 
\multicolumn{5}{l}{  \emph{Degradation process toward death before dementia} }\\
   Threshold        & Intercept       & 1.744   & 0.108    & $<$0.001 & 1.725  & 0.105 & $<$0.001 & 1.743  & 0.101 & $<$0.001 & 1.810  & 0.099  &$<$0.001\\ 
                    & S1(age)$^\dag$  & -0.956  & 0.201    & $<$0.001 & -1.204 & 0.206 & $<$0.001 & -1.002 & 0.199 & $<$0.001 & -1.335 & 0.186  &$<$0.001\\ 
                    & S2(age)$^\dag$  & -3.125  & 0.363    & $<$0.001 & -3.402 & 0.503 & $<$0.001 & -3.169 & 0.354 & $<$0.001 & -3.697 & 0.347  &$<$0.001\\ 
                    & S3(age)$^\dag$  & -4.150  & 0.414    & $<$0.001 & -4.440 & 0.626 & $<$0.001 & -4.233 & 0.411  & $<$0.001& -4.672 & 0.404  &$<$0.001\\ 
  Cognition         & Intercept       & -0.011  & 0.042    & 0.801    & 0.079  & 0.031 & 0.011    &        &       &        &        &         \\ 
  Dependency        & Intercept       & 0.178   & 0.045    & $<$0.001 &        &       &          & 0.108  & 0.032 &0.001   &        &         \\ 
  Dep. Sympt.       & Intercept       & -0.105  & 0.059    & 0.076    &        &       &          &        &       && 0.010  & 0.046  &   0.828   \\ 
\hline
\multicolumn{2}{l}{conditional AIC$^\ddag$} & \multicolumn{3}{r}{2021.0} & \multicolumn{3}{r}{2155.6 } & \multicolumn{3}{r}{2161.2} & \multicolumn{3}{r}{2438.1}\\
\multicolumn{2}{l}{conditional BIC$^\ddag$} & \multicolumn{3}{r}{2099.9} & \multicolumn{3}{r}{2208.2}  & \multicolumn{3}{r}{2218.3} & \multicolumn{3}{r}{2495.1}\\
\multicolumn{2}{l}{UACV$^\ddag$}            & \multicolumn{3}{r}{1.710}  & \multicolumn{3}{r}{1.825}   & \multicolumn{3}{r}{1.824}  & \multicolumn{3}{r}{2.055}\\
\hline
\end{tabular}
}
\newline
\begin{flushleft}
$^*$ Initial age is indicated in decades and centered around 65 years old.\\
\rev{$^\dag$ S1, S2, S3 are natural cubic spline functions applied on age}\\
\rev{$^\ddag$ The lower the better. Are reported Akaike and Bayesian Information Criteria (AIC, BIC) and Universal Approximate Cross Validation criterion (UACV) for dementia and death data conditional on longitudinal information. UACV differences of order 10$^{-1}$ , 10$^{-2}$ and 10$^{-3}$ qualified as "large", "moderate" and "small", respectively (\citealp{commenges2015}).}
\end{flushleft}
\end{table}

Estimates of the submodel for the degradation process toward dementia are given in Table \ref{tab_estimates2}. 
\rev{Adjusted for covariates, cognitive and functional domains significantly contributed to the degradation process toward dementia with increased impairments associated to higher degradation of the dementia process and a higher weight of function among ApoE4 carriers. }Interestingly, depressive symptoms did not contribute at all to the degradation process toward dementia among individuals with a high educational level ($\theta$=-0.447+0.452=0.005, p=0.957), but it highly contributed among those with a low educational level ($\theta$=-0.447, p=002); in this group, higher depressive symptoms induced a lower level of the degradation process for the same level of cognition and function. This result may be explained by the cognitive reserve linked with education. Among individuals who did not reach a sufficient educational level, having higher depressive symptoms alters the cognitive evaluation while individuals who reached a sufficient educational level are able to compensate and properly pass the cognitive evaluation. Finally, the threshold at which dementia is diagnosed differed according to covariates, with dementia diagnosed at a lower level of degradation for those with a higher educational level, a younger age or a diagnosis made after the 10year visit. No significant difference was found according to gender or ApoE4 status. 

\begin{figure}[!p]
\centering\includegraphics[width=1.0\textwidth]{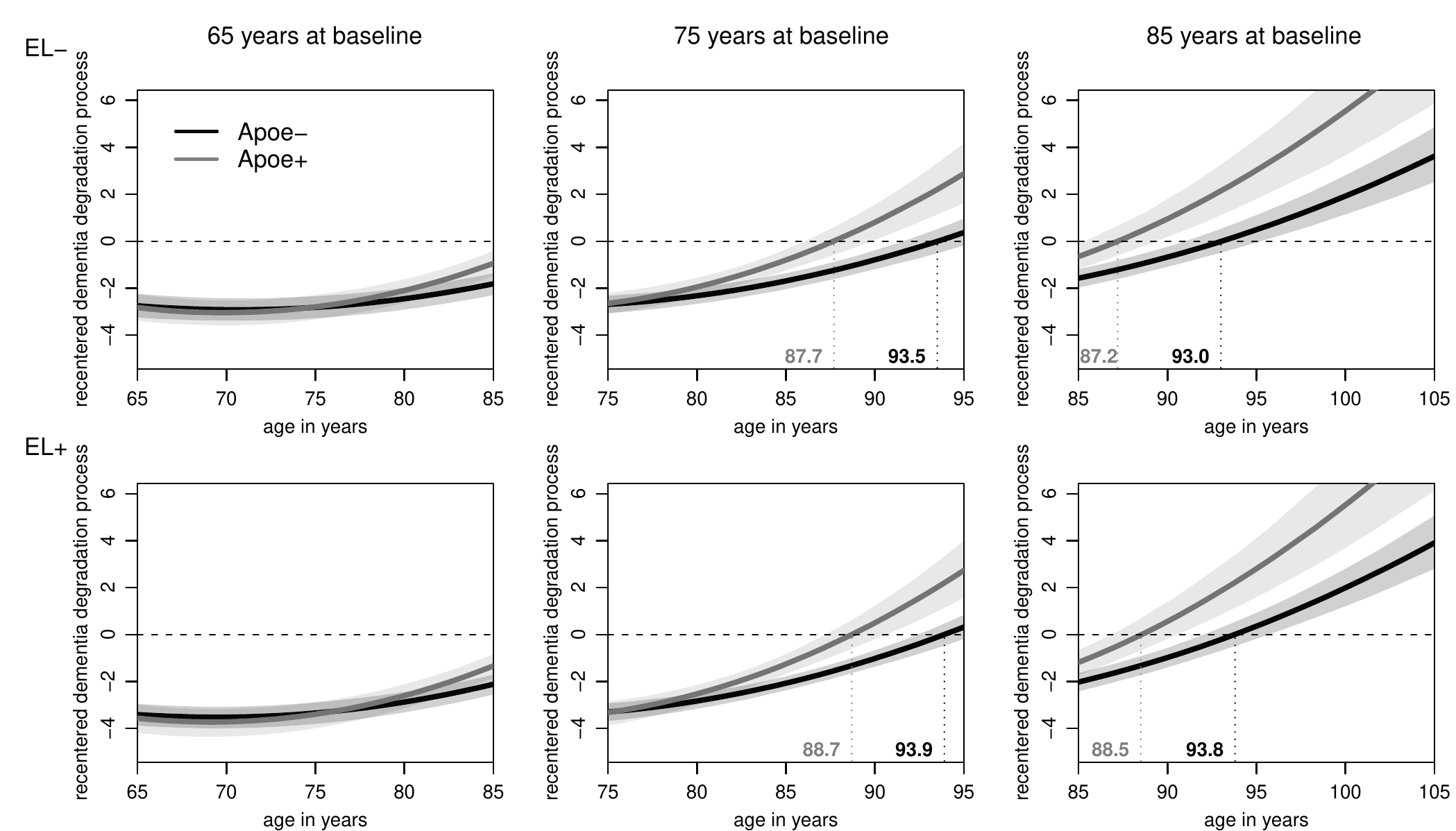}
\caption{Predicted degradation process toward dementia according to education level (EL+ top panel; EL- bottom panel), Initial age (65y left, 75y middle, 85y right) and ApoE4 status (non carrier in black, carrier in grey) with 95\% confidence bands computed by Monte Carlo (with 2000 draws). The degradation process was recentered so that 0 represents the threshold above which dementia diagnosis becomes positive for each profile.}
\label{fig_pred_deg}
\end{figure}

To illustrate these differences in the structure of the degradation process, Figure \ref{fig_pred_deg} displays the mean predicted trajectories of the degradation process according to education, age at entry and ApoE4 status. Note that here, the degradation process was recentered by absorbing the covariate-specific threshold so that dementia is diagnosed when the process is above 0. ApoE4 status constitutes the main modulating factor of the trajectory toward dementia with ApoE4 carriers diagnosed with a dementia 5 to 6 years before ApoE4 non carriers. In contrast, despite a different structure of the degradation process according to education, very limited differences were found which highlights that the differential contribution of depressive symptoms mainly served as a compensating factor in the degradation process definition.

\rev{We investigated the added value of simultaneously considering multiple domains in association with dementia and death rather than considering each domain separately. As longitudinal information differed, we focused on information criteria for dementia and death conditional to longitudinal information with conditional Akaike and Bayesian Information Criteria (AIC, BIC) (\citealp{zhang2014}) and Universal Approximate Cross Validation criterion (UACV) (\citealp{commenges2015}) (See calculations in supplementary Section 3.1). All criteria concluded to the large superiority of the multivariate model over univariate models with a difference of at least 0.1 in UACV, 130 points in AIC and 109 points in BIC (Table  \ref{tab_estimates2}). From an epidemiologic perspective, conclusions also differed. In univariate models, each domain contributed positively to the degradation process toward dementia even depressive symptomatology although, adjusted for cognition and function, depression was no more positively associated with the degradation process but contributed only among low educated people and negatively as a compensatory factor.}

\subsection{Supplementary results and analyses}

The estimated splines transformations linking each marker to its underlying domain (Figure S2 in Supplementary Materials) exhibited clear nonlinear relationships except for IST and BVRT cognitive tests. This illustrates the varying sensitivity to change of scales observed in previous works (\citealp{proust-lima_misuse_2011}) and the relevance of taking such nonlinear relationships into account. The correlation matrix between the 9 individual random effects also exhibited very high correlations between the level at 65 years, age and age$^2$ within and between cognitive, functional and depression domains with correlations up to 0.90 between slopes of cognitive and functional domains (Figure S3 in Supplementary Materials). \rev{In a secondary analysis, we considered the domains as independent. Although epidemiological results regarding the degradation processes toward death and dementia did not change, the fit of dementia and death data was substantially impacted (conditional AIC=2021.0, conditional BIC=2099.9, UACV=1.710). }

The goodness-of-fit of the model was check by verifying that the individual predictions were closed enough to the observations in the multivariate mixed submodel (Figure S4 in Supplementary Materials). We also verified that the model for dementia-free death considering thresholds approximated by splines was flexible enough by comparing it with a model considering interval-specific probability of death (Figure S5 in Supplementary Materials). 

\vspace*{-0.5cm}
\section{Discussion}
\label{sec-conclu}
We proposed a novel joint model for multiple longitudinal dimensions and clinical endpoints. Initially motived by the study of one repeated binary clinical endpoint measured at predefined visits, the model also applies to continuous endpoints such as death (provided they can be discretized) and to competing risk setting as shown in the application. 

The complexity in joint models, as induced for instance by multiple longitudinal markers, usually orientates the model development toward Bayesian approaches which may better tackle numerical problems. An originality of this work is that the estimation in the frequentist framework was made possible thanks to properties of skew-normal distributions which avoided the cumbersome multiple numerical integration over the random effects usually encountered in joint shared random effect models (\citealp{rizopoulos_fully_2009}). As such, this approach can be applied to contexts where the number of random effects becomes substantial (such as 9 in the application) and/or correlated Gaussian processes are added to relax the model. Being able to jointly model a substantial number of markers becomes indeed a real challenge with the growing availability of dynamic markers in longitudinal health studies.

The properties of skew-normal distributions had already been exploited to simplify inference in joint models in the case of a single repeated biomarker and a single continuous time to event (\citealp{barrett_joint_2015}). The authors had opted for a sequential probit model for the discretized time where we opted for a degradation process model (Section \ref{sec_degrad_death}). Although different in their definition, the two approaches are numerically equivalent in the absence of delayed entry as shown in Section 1.3 of Supplementary Materials. 

In Alzheimer's disease, this model gives for the first time an opportunity to describe the multidimensional pathological process toward dementia. We considered cognitive, functional and depressive symptomatology measures to understand how they contributed to the degradation process toward dementia. \rev{Taken independently, each component was associated with the degradation toward dementia. However, when considered jointly, we found an interesting compensatory mechanism of depressive symptomatology among lower educated participants.} Depressive symptomatology seems to counterbalance their cognitive evaluation which does not correctly translate their actual cognitive level in the case of depressive symptomatology, probably due to a poorer cognitive reserve. This mechanism impacted only the definition of the dementia degradation process, not its level. We also confirmed and quantified the higher propensity of ApoE4 carriers to develop dementia. Next step will be to also consider neurodegeneration information to capture the anatomic component of dementia (\citealp{jack_tracking_2013}).

The approach however has several limits. Although elegant, the technique is limited to manageable dimensions of repeated clinical endpoints at the individual level. The likelihood calculation relies on algorithms to compute the multivariate normal cumulative distribution function. Their good precision may become questionable beyond very large dimensions which probably limits the methodology to no more than two competing events. Such limitation remains at the individual level, not at the population level which may still include a lot more intervals and/or diagnosis visits. In addition, we believe this is a reasonable concession in many applications where multivariate longitudinal dimensions with a large amount of random effects are to be modelled and could not by using the numerical approaches previously proposed in the literature. Another limitation is that the methodology does not apply to repeated binary, ordinal or count markers. However, by using parameterized link functions, we can handle continuous non Gaussian markers which permits to correctly analyze many asymmetric scales (\citealp{proust-lima_misuse_2011}). 
For the latent process models, we chose to assume that one marker was the manifestation of only one latent process as usual in latent variable methodology and that the possible correlation between the markers was handled at the latent processes level. This was relevant in Alzheimer's disease where domains under study are defined from distinct families of markers. However, an interesting alternative may be to consider independent principal latent processes built from all the markers in the spirit of principal component analyses. Finally, to adapt to the context of Alzheimer's disease, we defined latent processes measured by multiple markers which may not be necessary in other contexts where markers directly constitute the components under study. \rev{The methodology and the program made available under R apply in such a case as the standard multivariate linear mixed model constitutes a specific case of the approach.}

To conclude, by handling multivariate repeated markers and clinical endpoint, this joint model opens to many applications in which identifying markers of clinical progression are of importance. The definition of the clinical endpoint differs from most joint model proposals but it is actually clinically relevant in many diseases characterized, as in neurodegenerative diseases, by a body of progressive impairments. 

\vspace*{1cm}

\subsection*{Acknowledgements}
This work was supported by project SMALA (ANR-15-CE37-0002) of the French National Research Agency (ANR) and project 2CFaAL of France Alzheimer Association. Computer time was provided by the computing facilities MCIA (M\'esocentre de Calcul Intensif Aquitain) at the Universit\'e de Bordeaux and the Universit\'e de Pau et des Pays de l'Adour.

\subsection*{Data Availability Statement}
The data that support the findings of this study are available from the corresponding author upon reasonable request with exception to the application raw data from PAQUID cohort. The software is openly available in github at https://github.com/VivianePhilipps/multLPM.

\bibliographystyle{biom}
\bibliography{refs}

\appendix

We used the following equality obtained by \cite{arnold_flexible_2009} (equation 59 of the paper) for skew-normal variables to compute an exact formula of the joint model likelihood:  

\begin{equation}
\Phi^{(m)} \left ( \lambda_0 ; 0, \Delta + \Lambda \Lambda^T \right ) = \int_{\mathbb{R}^{k} } \Phi^{(m)}\left ( \lambda_0 + \Lambda z ; 0, \Delta \right ) \phi^{(k)} (z) dz
\end{equation}
where $\lambda_0$ is a m-vector, $\Delta$ is a $ m\times m$ variance covariance matrix and $\Lambda$ is a $m\times k$ matrix. As in the main text, $\Phi^{(n)}(\boldsymbol{x};\boldsymbol{m},\boldsymbol{V})$ denotes the $n$-dimensional Gaussian cumulative distribution function with mean $\boldsymbol{m}$ and variance $\boldsymbol{V}$ computed in $\boldsymbol{x}$, and $\phi^{(k)}$ is the $k$-dimensional standard Gaussian density function.

\end{document}